# Efficiency analysis of betavoltaic elements

A.V. Sachenko[1*], A.I. Shkrebtii[2], R.M. Korkishko[1], V.P. Kostylyov[1], M.R. Kulish[1], I.O. Sokolovskyi[1]

[1] V. Lashkaryov Institute of Semiconductor Physics, NAS of Ukraine,
41 prospect Nauky, 03028 Kyiv, Ukraine
[2] University of Ontario Institute of Technology, Oshawa, ON, Canada
* Corresponding author. E-mail: sach@isp.kiev.ua. Phone: +38(044)525-57-34

Abstract

The conversion of energy of electrons produced by a radioactive $\beta$-source into electricity in a Si and SiC $p-n$ junctions is modeled. The features of the generation function that describes the electron-hole pair production by an electron flux and the emergence of a "dead layer" are discussed. The collection efficiency $Q$ that describes the rate of electron-hole pair production by incident beta particles, is calculated taking into account the presence of the dead layer. It is shown that in the case of high-grade Si $p-n$ junctions, the collection efficiency of electron-hole pairs created by a high-energy electrons flux (such as, e.g., Pm-147 beta flux) is close or equal to unity in a wide range of electron energies. For SiC $p-n$ junctions, $Q$ is near unity only for electrons with relatively low energies of about 5 keV (produced, e.g., by a tritium source) and decreases rapidly with further increase of electron energy. The conditions, under which the influence of the dead layer on the collection efficiency is negligible, are determined. The open-circuit voltage is calculated for realistic values of the minority carriers' diffusion coefficients and lifetimes in Si and SiC $p-n$ junctions, irradiated by a high-energy electrons flux. Our calculations allow to estimate the attainable efficiency of betavoltaic elements.

## *1. Introduction*

Betavoltaic effect refers to conversion of energy from electrons generated in nuclear reactions into electricity in semiconductor $p-n$ junctions. Long action periods (decades) and the ability to operate in a wide temperature range (from –50 to 150º C) make betavoltaic effect attractive for a number of technological applications, such as communication devices, sensors in hard-to-reach areas, and implantable medical devices [1]. In particular, a battery for heart pacemakers based on the Si $p-n$ junction and Pm-147 source of beta particles was produced [2]. Other popular beta-source and recipient are tritium and $A_3B_5$ semiconductors [3-5].

After the pioneering work of Ehrenberg, et al. [6], betavoltaic conversion has been intensively investigated both experimentally [7-10] and theoretically [8], [9]. Of particular importance in this respect is the efficiency of a betavoltaic cell, defined as the ratio of the power generated by the cell to the power delivered by the incident electrons. The first estimates of this parameter were performed in Refs. [2, 10-12].

Betavoltaic and a more familiar photovoltaic effect are related and share several traits. In both cases, calculations of the limiting efficiency are made under the assumption that all electron-hole pairs produced by the incident beta particles or photons contribute to the betavoltaic or photovoltaic current. When viewed as a function of the band gap, $E_g$, the current generated by beta-particles decreases, whereas the open-circuit voltage grows with $E_g$. Because the latter effect dominates over the former, the overall betaconversion efficiency increases with $E_g$ [2]. In the case of photovoltaics, in contrast, the photocurrent decrease with band gap is stronger, resulting in a maximum of photoconversion efficiency as a function of $E_g$ [13].

In this research the attainable betavoltaic conversion efficiency is modeled using realistic values of minority carriers' lifetimes and diffusion coefficients to calculate the collection coefficient $Q$ and open circuit voltage $V_{OC}$. The dominating recombination mechanisms were included in the formalism developed. The results were compared to the limiting photoconversion efficiency, which corresponds to the fundamental maximum of $\eta$.

For the betavoltaic cell we assume that the $\beta$ source is in the form of a foil [2]. This allows considering the one-dimensional (slab) geometry for our theoretical analysis. The cross-section of the betavoltaics cell and $\beta$-source is shown schematically in Fig. 1. The beta electrons flux is directed toward $p - n$ junction, which separates generated electron-hole pairs, are shown. The dead layer, where the scattering the beta-electrons can be neglected, is $x_m$ thick, and is located close to the frontal surface of the sample and the $\beta$ source.

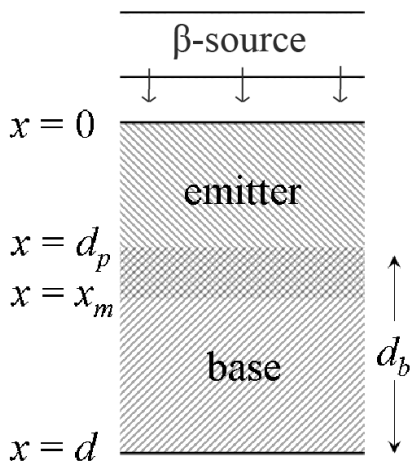

Fig. 1. Cross-section of the betavoltaic system with *β*-source adjacent to *p – n* junction. The emitter and the base are $d_p$ and $d_b$ thick, and *p – n* junction width $d = d_p + d_b$. The dead layer is $x_m$ thick.

In the beta battery, each beta-electron produces a number of electron-hole pairs by dissipating its energy *E* in the semiconductor. Electron energy is in turn averaged with respect to the decay spectrum for each beta emitter [14]. Experimental spectra for tritium from [15] and promethium from [16] were used in beta-conversion efficiency calculation. Electron-hole pair generation function by a beta electron can be written as [17].

## 2. The collection efficiency analysis

Efficiency $\eta$ of a betavoltaic converter can be written as a product of three terms [2]

$$\eta = \eta_\beta \eta_c \eta_s . \tag{1}$$

Here

$$\eta_\beta = N_\beta / N_0 \tag{2}$$

is the ratio of the beta-flux $N_\beta$, reaching the semiconductor surface, to the total flux $N_0$ emitted;

$$\eta_c = (1-r)Q \tag{3}$$

is the coupling efficiency, given by the product of absorption probability of a beta-particle (*r* is the coefficient of electron reflection from the semiconductor surface) and collection efficiency *Q* of electron-hole pairs; finally, the semiconductor efficiency is

$$\eta_s = qV_{OC} FF / \varepsilon . \tag{4}$$

Here *q* is the elementary charge, $V_{OC}$ is the open circuit voltage, *FF* is the fill factor of the current-voltage characteristics, $\varepsilon = (2.8 E_g + 0.5)$ eV is the energy needed to create one electron-hole pair, and $E_g$ is semiconductor band gap [18].

Previously, the maximal betavoltaic efficiency was obtained under an implicit assumption that each incident beta-particle produces electron-hole pairs, i.e., the collection efficiency *Q* = 1. To derive the collection efficiency *Q*, we initially also used the expression $g(x) \propto \exp(-\alpha x)$ for the generation function for electron-hole pairs created by beta-electron flux (see [8, 11]), where $\alpha$ is the absorption coefficient, *g* is the generation rate of electron-hole pairs per unit volume, and *x* is the distance from the semiconductor front surface. This expression, however, is valid only outside of the dead layer when $x > x_m$, while $g(x)$ is assumed be zero at $x < x_m$ [19]. Sufficiently general expressions for $Q$ in this approximation are given in [8, 11], while the most general one is derived in [20]. It has the following form:

$$Q = Q_p + Q_n, \tag{5}$$

$$Q_p = \frac{\alpha L_p}{(\alpha L_p)^2 - 1} \cdot \frac{\alpha L_p + S_0 \frac{\tau_p}{L_p}\left(1 - e^{-\alpha d_p}\right)\cosh\left(\frac{d_p}{L_p}\right) - e^{-\alpha d_p}\sinh\left(\frac{d_p}{L_p}\right) - \alpha L_p e^{-\alpha d_p}}{S_0 \frac{\tau_p}{L_p}\sinh\left(\frac{d_p}{L_p}\right) + \cosh\left(\frac{d_p}{L_p}\right)}, \tag{6}$$

$$Q_n = \frac{\alpha L e^{-\alpha d_p}}{1-(\alpha L)^2} \cdot \left\{ \frac{\left[S_d \cosh\left(\frac{d}{L}\right) + \frac{D}{L}\sinh\left(\frac{d}{L}\right)\right]\left(1 + r_d e^{-2\alpha d}\right) + \left(\alpha D(1-r_d) - S_d(1+r_d)\right)e^{-\alpha d}}{S_d \sinh\left(\frac{d}{L}\right) + \frac{D}{L}\cosh\left(\frac{d}{L}\right)} \right.$$

$$\left. - \frac{\alpha L S_d \left[\sinh\left(\frac{d}{L}\right) + \frac{D}{L}\cosh\left(\frac{d}{L}\right)\right]\left(1 - r_d e^{-2\alpha d}\right)}{S_d \sinh\left(\frac{d}{L}\right) + \frac{D}{L}\cosh\left(\frac{d}{L}\right)} \right\}. \tag{7}$$

Here, $Q_p$ and $Q_n$ are the collection coefficients of electron-hole pairs in the emitter and the base respectively, $L_p$ is the diffusion length in the emitter, $d_p$ is the thickness of the emitter, $\tau_p$ is the bulk lifetime in the emitter, $r_d$ is the electron reflection coefficient from the back surface of the $p-n$ junction, $L = (D\tau_b)^{1/2}$ is the diffusion length in the base, $D$ and $\tau_b$ are the base diffusion coefficient and the bulk lifetime, respectively, $d$ is the thickness of the base, $S_0$ is the effective surface recombination rate at the emitter surface, and $S_d$ is the effective surface recombination rate at the back surface of the base.

In the case of a beta battery, each beta-electron dissipating its energy $E$ in the semiconductor produces a number of electron-hole pairs. Electron energy is in turn averaged with respect to the decay spectrum in beta-conversion efficiency calculation. Electron-hole pair generation function by a beta electron can be written as [17]

$$g(x) = -J(x)\frac{1}{\varepsilon}\frac{dE}{dx} = J(x)\frac{1}{\varepsilon}\frac{2\pi q^4 NZ}{E(x)}B[E(x)]. \tag{8}$$

Here $J(x)$ is the density of electrons flux, $(-\frac{dE}{dx})$ is the energy dissipation per unit path length, $N$ is the number of absorber atoms per cm$^3$, $Z$ is the atomic number of the absorber material, and $B(E)$ is the stopping number. The electron energy dissipation per unit path length can be calculated utilizing (8). The transcendent solution of (8) multiplied by the electron energy distribution function and integrated over the energy is fit well by the exponential generation function in the form $\frac{E_0}{\varepsilon}\alpha\exp(-\alpha x)$ [21].

The factor $\alpha$ in (5) - (7) describes the decay of excess concentration of electron-hole pairs generated by beta electrons. Comparing the averaged by decay energy left part of (8) to the exponential generation function, one can determine the $\alpha$-value. This procedure yields $\alpha = 300$ cm$^{-1}$ for Pm-147/Si, 270 cm$^{-1}$ for Pm-147/SiC, 3·10$^4$ cm$^{-1}$ for T/Si, and 1.5·10$^4$ cm$^{-1}$ for T/SiC.

The collection efficiency of a $p-n$ junction based on high-quality silicon with Shockley-Read-Hall recombination lifetime $\tau_{SR} \approx 1$ ms can be calculated using (5) - (7). Due to the low currents in betavoltaic elements, the excess concentration of electron-hole pairs, $\Delta p$, is much smaller than the equilibrium majority charge carriers (holes) concentration in the base even for very long lifetimes. Considering the Shockley-Read-Hall lifetime $\tau_{SR}$ in the base, radiative recombination coefficient $B_r$, and Auger interband recombination coefficient $C_{Auger}$, the resultant bulk carrier lifetime $\tau_b$ can be written as

$$\tau_b = \left(\tau_{SR}^{-1} + B_r p_0 + C_{Auger} p_0^2\right)^{-1}, \tag{9}$$

where $p_0$ is the equilibrium majority carriers' concentration in the base.

Figure 2 (a) shows the collection coefficient $Q$ decrease with the increase in the effective surface recombination rate $S_d$ for silicon $p-n$ junctions. In this case, the mean energy of Pm-147 electrons is 61.9 keV. The parameters used to calculate Si and SiC collection coefficients are given in Table 1. Absorption coefficient $\alpha$ is taken from (8), using the approach of [17]. For the above considered parameters, the collection coefficient depends very weakly on the recombination rate $S_d$ and always exceeds the rather high value of 0.85. The parameters of beta-sources are given in Table 2.

For tritium as the source of beta-electrons with the mean energy of 5.7 keV, $\alpha$ is assumed to be 1,5·10$^4$ cm$^{-1}$ for SiC [17], and $Q$ for this case very close to unity (see Fig. 2(b)). Since the holes lifetime is higher than the one for electrons in SiC [22], *n*-type base was considered in this case to ensure large diffusion length. Standard *p*-type base was used in Si.

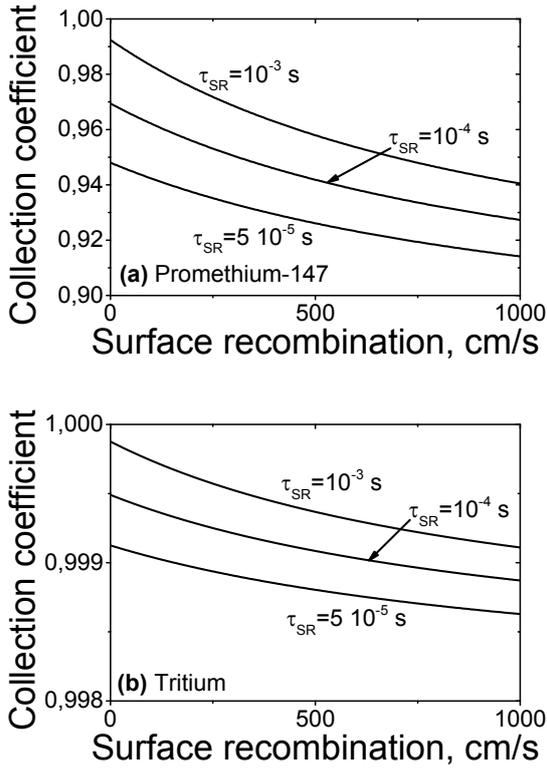

Fig. 2. The collection coefficient $Q$ vs. the effective surface recombination velocity for silicon $p-n$ junctions with Pm-147 (a) and tritium (b) as the electrons source.

|  | $L_p$, cm | $\tau_p$, s | $d_p$, s | $p_0$, cm$^3$ | $d$, cm | $D$, cm$^2$/s | $r_d$ | $B_r$, cm$^3$/s | $C_{Auger}$, cm$^6$/s |
|---|---|---|---|---|---|---|---|---|---|
| Si | $10^{-3}$ | $10^{-7}$ | $10^{-5}$ | $10^{17}$ | $3\cdot10^{-2}$ | 30 | 0.6 | $6\ 10^{-15}$ | $10^{-31}$ |
| SiC | $6\cdot10^{-6}$ | $10^{-10}$ | $10^{-6}$ | $10^{17}$ | $3\cdot10^{-2}$ | 1 | 0.6 | $1.5\cdot10^{-12}$ | $2\ 10^{-31}$ |

Table 1: Set of the semiconductor parameter used in (5)-(7).

| Source | α (Si), cm$^{-1}$ | α (SiC), cm$^{-1}$ | Mean energy, keV | Maximal Energy, keV |
|---|---|---|---|---|
| Pm-147 | 300 | 270 | 61.9 | 224,6 |
| T | 30 000 | 15 000 | 5.7 | 18,6 |

Table 2: The absorption coefficients α for different combinations of beta-source and semiconductors used in the calculations, and the mean and the maximum energies of beta-particles from the Pm-147 and T sources.

Figure 3 (a) shows the collection coefficient $Q$ as a function of the effective surface recombination velocity $S_d$ for SiC $p-n$ junctions. The dependences were calculated using the parameters from Tables 1 and 2. The figures demonstrate that the collection coefficients $Q$ for this case are small compared to unity and do not depend on $S_d$, because the $L \ll d$ condition is satisfied. This criterion means high losses of minority carriers in the bulk. Therefore, the attainable efficiency of the SiC betavoltaic battery with $Pm-147$ is smaller than the efficiency of the high-quality silicon battery.

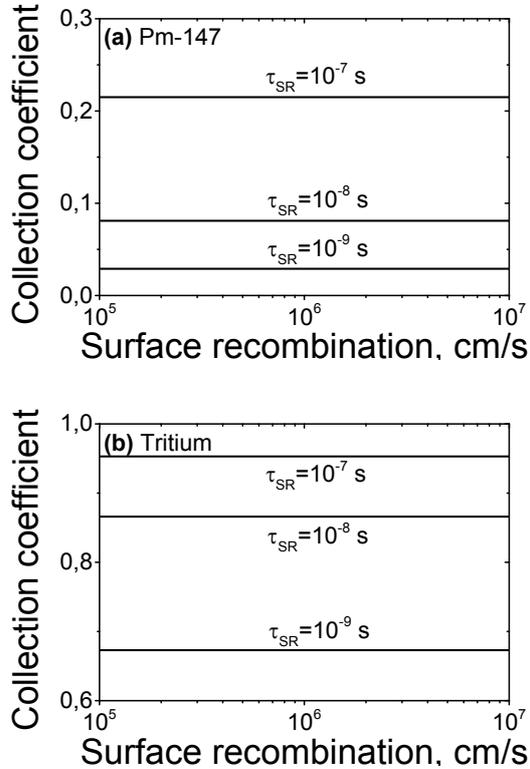

Fig. 3. The collection coefficient $Q$ vs. the effective surface recombination velocity for SiC $p-n$ junctions with Pm-147 (a) and tritium (b) as the electrons source.

The graphs in Fig. 3 (b) were calculated for tritium as the source of beta-electrons and using the effective surface recombination velocity $S_d$. Other parameters are the same as for Fig. 3 (a). Collection coefficient $Q$ in this case is practically independent from $S_d$, being rather high even for short Shockley-Read-Hall recombination time $\tau_{SR} = 10^{-9}$ s, resulting in $Q = 0.67$. This value of collection coefficient $Q$ is acceptable for practical applications. Therefore, SiC battery efficiency increase should be expected if tritium source is used. The $\alpha L \geq 1$ criterion ensures the effective electron-hole pairs collection in this case. In the case of silicon, this criterion is well satisfied for both Pm-147 and for tritium sources. However for SiC, $\alpha L$ is close to 1 only for tritium source.

In the beginning of this section we initially wrote the electron-holes generation function as $g(x) \propto \exp(-\alpha x)$. Since according to [8], the generation function has a maximum at $x_m$, the

exp($-\alpha x$) form is correct for distances exceeding $x_m$. This maximum is due to negligible electron scattering until $x \geq x_m$. For gallium arsenide under beta irradiation, $x_m$ is in the range of $10^{-5} \div 10^{-4}$ cm [19]. The greater the electron energy, the greater $x_m$ is. The area of $x < x_m$, is termed the dead layer [19]. The collection efficiency can be significantly reduced for the case $L < x_m$. To illustrate this, an expression for the collection efficiency $Q$ was derived under the assumption that the electron-hole pair generation is absent for $x < x_m$ and is described as $g(x) = I_0 \exp(-\alpha x)$ for $x \geq x_m$. We also assume that $d_p < x_m$, $\alpha d > 1$ and $L < d/2$. Then, solving the continuity equation for the excess electron-hole pairs concentration $\Delta p$ in the region 1 (where $x \leq x_m$) and in the region 2 (where $x \geq x_m$) with the boundary conditions $\Delta p_1(x = d_p) = 0$, $\Delta p_2(x = d) = 0$ and matching the solutions, we find the integration constants. Collection efficiency is then determined from the expression $Q = -(D \cdot d\Delta p_1 / dx)/I_0$ at the point $x = d_p$ and it has the following form:

$$Q \cong \frac{\alpha L}{1 + \alpha L} \exp\left(-\frac{x_m - d_p}{L}\right). \tag{10}$$

Figure 4 shows the dependence of $Q$ on the diffusion length, calculated by (10). The depth of the $p-n$ junction was fixed ($d_p = 10^{-5}$ cm), while the beta-electron energy (and so the absorption coefficient $\alpha$) and $x_m$ were varied. The dashed curves are plotted using $Q = \alpha L/(1+\alpha L)$ expression, which becomes correct in the absence of the dead layer.

Figure 4(a) shows that biggest collection efficiency reduction occurs for the upper curves with tritium as an electron source. The discrepancy between $Q$ values calculated with and without taking the dead layer into account is minimal for lowest curves, for which $Pm-147$ is a source of electrons. In this case, to obtain sufficiently large collection efficiency $Q > 1/2$, the diffusion length $L$ must exceed 35 μm. High $Q$ values ($Q \sim 1$) can only be achieved in silicon $p-n$ junctions with large minority charge carriers lifetime.

The graphs in Fig 4 (b) are plotted for $x_m = 10^{-5}$ cm and $\alpha = 4 \cdot 10^4$ cm, which corresponds to the tritium decay. The figure demonstrates that the $Q$ increases with $L$ and when the junction depth is approaching $x_m$. This effect is particularly significant for small diffusion lengths. Thus, the analysis shows that diffusion length $L$ higher than $x_m$ is needed for sufficiently high electron-hole pairs collection efficiency. An alternative way to increase $Q$ is to use a deep $p-n$ junction with $d_p \approx x_m$.

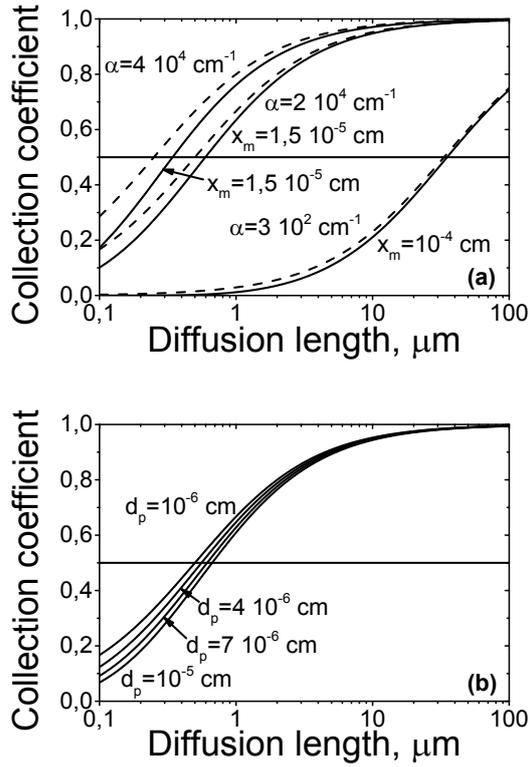

Fig. 4. Collection efficiency $Q$ dependence on the minority carriers' diffusion length in the base $L$ for the SiC $p-n$ junction, calculated taking into account existence of the dead layer (solid lines). Pm-147 is the beta-electron source in (a), while the tritium source is considered in (b).

The results of $Q$ calculations without considering the dead layer shown in Figs. 2 and 3 are accurate enough for $d_p = x_m$. Indeed, Fig. 4(a) shows that $Q$ does not change substantially for $d_p < x_m$ in SiC $p-n$ junctions. With SiC and tritium, $Q$ from Fig. 3(b) is reduced by a factor of $\exp[-(x_m - d_p)/L]$ for $d_p < x_m$. Therefore, the $Q$ value is 0.83 for the base lifetime of $10^{-8}$ s, $x_m = 2 \cdot 10^{-5}$ cm, $d_p = 10^{-6}$ cm and $L = 10^{-4}$ cm. This reduction does not lead to drastic changes of the $Q$ values shown in Fig. 3(b).

## 3. Open circuit voltage analysis

It should be noted that the calculated in [2,11] open circuit voltage, collection efficiency, and therefore the limiting efficiency as a function of band gap were overestimated. However, it will be shown below that the efficiency overestimation is particularly high for wide-gap semiconductors having much lower bulk lifetimes than for high-quality silicon. For two-component wide-gap semiconductors a significant degree of compensation is also common, which decreases $V_{OC}$.

To find the open-circuit voltage $V_{OC}$ in the realistic case (with the system parameters close to the experimentally achievable) and taking into account such recombination mechanisms as recombination in the quasi-neutral region and recombination in the space charge region (SCR) with the width $w$, the generation-recombination balance equation should be written as:

$$J_\beta = q\left[\frac{D}{L}\tanh\left(\frac{d}{L}\right) + S_0\right]\Delta p^* + qV_{SC}\Delta p^* . \qquad (11)$$

Here $J_\beta \cong 3.64 J_0 Q/(2E_g + 0.5)$ is the density of the beta-electron-excited current, $J_0$ is the density of current generated by beta-particles in silicon, $S_0$ is the effective surface recombination velocity at the front surface of the cell, $\Delta p^* = \Delta p|_w$, $L = \sqrt{D\tau_b}$, and

$$V_{SC}(\Delta p) = \frac{L_D}{\sqrt{2}\tau_{SR}}\int_{y_{pn}}^{-1}\frac{N_d - N_a}{\left[(N_d - N_a)e^y + n_i e^{\varepsilon_r}) + b\left(\frac{n_i^2}{N_d - N_a} + \Delta p^*\right)e^{-y} + n_i e^{-\varepsilon_r}\right]F}dy$$

is recombination in the SCR. Here $F$ is the dimensionless electric field in the space-charge region,

$$F = \left\{-y + \frac{N_a}{N_d}\left[\ln\left(1 + \exp\left(-\varepsilon_a + y + \ln\left(\frac{N_d - N_a}{n_i}\right)\right)\right) - \ln\left(1 + \exp\left(-\varepsilon_a + \ln\left(\frac{N_d - N_a}{n_i}\right)\right)\right)\right] + \frac{N_d - N_a}{N_d}(e^y - 1)\right\}^{1/2}, \quad b = C_p/C_n.$$

$\varepsilon_a = E_a/kT$ and $\varepsilon_r = E_r/kT$ are dimensionless energies of the acceptor and recombination levels with respect to the middle of the band gap, $N_d$ and $N_a$ are donor and acceptor concentrations, $y$ is the dimensionless current potential, $y_{pn}$ is the dimensionless potential at the $p-n$ interface. $L$ and $D$ are the diffusion length and diffusion coefficient of the minority carriers, $\tau_{SR} = (C_p N_r)^{-1}$ is the Shockley-Read-Hall lifetime, $N_r$ is the concentration of deep recombination centers, $N_d - N_a$ is the equilibrium concentration of majority carriers in quasi-neutral region of the base, $\Delta p$ is the excess concentration of minority carriers in the base on the boundary between the space-charge region and the quasi-neutral region. The value of $J_0$ usually varies in the range of $1 \div 10^2$ μA / cm$^2$ [2].

Open circuit voltage is described by the relation

$$V_{OC} = \frac{kT}{q} \ln\left(\frac{n_0 \Delta p}{n_i^2}\right). \tag{12}$$

Here $n_i$ is the concentration of charge carriers in an intrinsic semiconductor. To calculate $V_{OC}$ the solution of (11) should be substituted in (12).

Fig. 5 shows the $V_{OC}$ dependence on the base doping level $p_0$ for silicon $p-n$ junction. Shockley-Read-Hall recombination, radiative recombination, and interband Auger recombination were taken into account in our calculations. No compensation was assumed, while the Shockley-Read-Hall lifetime values of $10^{-3}$, $10^{-4}$ and $5\cdot 10^{-5}$ s were used. Calculations were carried out for the temperature of 310 K, which corresponds to the human body temperature, and was taken as constant. One can see from the figure that $V_{OC}(p_0)$ has the maximum at high lifetimes, which disappears with lifetime decrease. The highest $V_{OC} = 0.58$ V is reached at $\tau_{SR} = 10^{-3}$ s and $p_0 = 10^{17}$ cm$^{-3}$. The highest $V_{OC}$ is reached at the current density of 100 µA / cm$^2$ for silicon. Such current density is more than two orders of magnitude smaller than the typical photocurrent in the silicon solar cells.

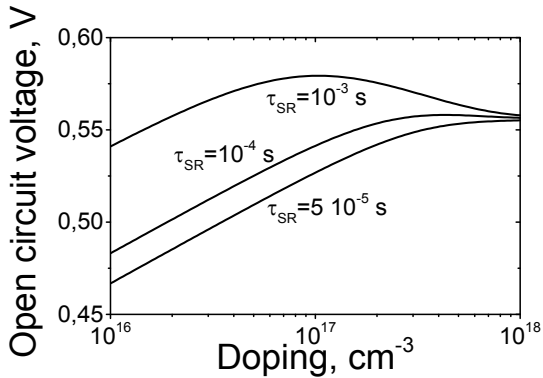

Fig. 5. Open circuit voltage *vs*. doping level for silicon $p-n$ junction. The following parameters were used: $J_0 = 100$ µA/ cm$^2$, $D = 30$ cm$^2$ /s, $T = 310$ K, $d = 300$ µm.

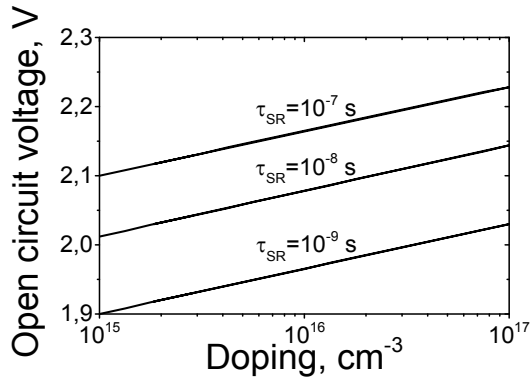

Fig. 6. The dependence of the open circuit voltage on doping level for SiC $p-n$ junction. The following parameters were used: $J_0 = 100$ µA /cm$^2$, $D = 1$ cm$^2$ / s, $T = 310$ K, $d = 300$ µm.

Fig. 6 shows the open-circuit voltage $V_{OC}$ vs. base doping level $N_d$ for SiC $p-n$ junction. Compensation was not taken into account in the calculation, i.e., $N_a$ value was assumed to be zero. Bulk lifetimes of $10^{-7}$, $10^{-8}$ and $10^{-9}$ s were used, $E_g$ value was assumed to be 3 eV. In contrast to Si, the $V_{OC}(N_d)$ curve increases logarithmically in this case. For the acceptor levels energy close to the valence band, the compensation was taken into account by using of the $V_{OC}$ value calculated for $n_0 = N_d - N_a$. Therefore, at high compensation level taking, e.g., $N_d = 10^{17}$ cm$^{-3}$ and $N_a = 9.99 \cdot 10^{16}$ cm$^{-3}$, $V_{OC}$ should be calculated for $n_0 = 10^{14}$ cm$^{-3}$. For example, open-circuit voltage $V_{OC} = 2.35$ V for the top curve is at $n_0 = 10^{17}$ cm$^{-3}$; for $n_0 = 10^{14}$ cm$^{-3}$ open circuit voltage $V_{OC} = 2.16$ V is significantly lower. The open circuit $V_{OC}$ value of 2.16 V is reached for $J_q = 1$ μA / cm$^2$ and $n_0 = 10^{17}$ cm$^{-3}$, while $V_{OC} = 1.96$ V for $n_0 = 10^{14}$ cm$^{-3}$.

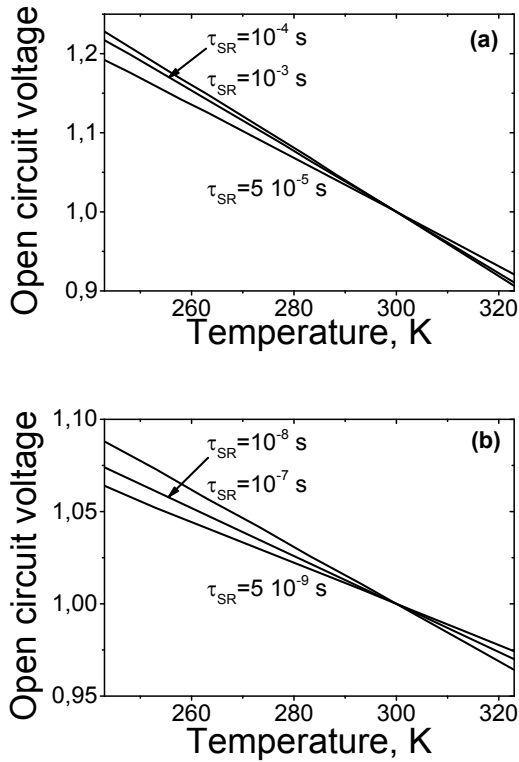

Fig. 7. The open circuit voltage $V_{OC}$ (normalized to its value at 300 K) vs. the external temperature. (a) Silicon $p-n$ junction. (b) SiC $p-n$ junction. Parameter used: $p_a = N_d = 10^{17}$ cm$^{-3}$, $J_q = 100$ μA /cm$^2$, $d = 300$ μm. The values of $D$ and $E_g$ in 7(a) are 30 cm$^2$/s and 1.12 eV; and in 7 (b) they are 1 cm$^2$/s and 3 eV.

When the external temperature is changing, $V_{OC}$ value decreases with temperature. Figure 7 shows the temperature dependence of $V_{OC}$ normalized to the value at 300 K ($v(T) = V_{OC}(T)/V_{OC}(300K)$). The temperature is varied in the range from -40 to 40º C. The temperature dependence of $V_{OC}$ is predicted to be stronger in the case of Si $p-n$ junctions than for SiC ones, because of the smaller band gap in Si. It should be noted that the slope of $V_{OC}(T)$ is defined by semiconductor band gap and minority carriers' lifetime. Fig. 7 shows that the higher minority carriers' lifetime the smaller $V_{OC}(T)$ slope.

Therefore, both $\tau_b$ value and the compensation factor affect $V_{OC}$ for the realistic parameters of the battery: the lower lifetime and the higher compensation, the lower $V_{OC}$.

The attainable conversion efficiency for the realistic systems can be written as

$$\eta_{real} = \eta_{\lim} Q \frac{V_{OC}}{V_{OC\lim}}, \tag{13}$$

where $\eta_{\lim}$ is the limiting efficiency and $V_{OC\lim}$ is the limiting open circuit voltage. The limiting efficiency $\eta_{\lim}$ was calculated as

$$\eta_{\lim} = \frac{qV_{OC\lim}FF_{\lim}}{2.8E_g + 0.5}. \tag{14}$$

Here $FF_{\lim}$ is the limiting fill factor defined as

$$FF_{\lim} = \left[\frac{V_{OC\lim}}{kT} - \ln(\frac{V_{OC\lim}}{kT} + 0.72)\right] / \left(\frac{V_{OC\lim}}{kT} + 1\right). \tag{15}$$

$\eta_{\lim}$, $V_{OC\lim}$ and $FF_{\lim}$ were calculated using the modified Shockley-Queisser approach [23] using $r = 0$, $Q = 1$ and $\eta_\beta = 1$. Figure 8 shows the limiting efficiency $\eta_{lim}$ vs. bandgap $E_g$. The efficiency was calculated by (14) using silicon radiative recombination parameter and effective densities of states for -arbitrary bandgap. The similar dependence from [2] (Fig. 4) is also included for comparison.

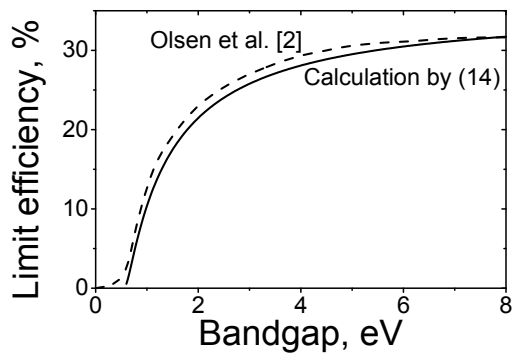

Fig. 8. The limiting betaconversion efficiency $\eta_{lim}$ calculated by (14) vs. semiconductor bandgap $E_g$, solid line. The dashed line shows the same dependence as calculated in [2].

| Semiconductor /emitter | $V_{OClim}$, V | $V_{OCreal}$, V | $\eta_{lim}$, % | $\eta_{real}$, % ($\tau_{SR}$, s) | $\eta_{real}$, % ($\tau_{SR}$, s) | $\eta_{real}$, % ($\tau_{SR}$, s) |
|---|---|---|---|---|---|---|
| Si/T | 0.53 | 0.48 | 11.8 | 10.7 ($10^{-3}$) | 9.8 ($10^{-4}$) | 9.6 (5 $10^{-5}$) |
| Si/Pm-147 | 0.65 | 0.60 | 14.9 | 13.6 ($10^{-3}$) | 12.4 ($10^{-4}$) | 12 (5 $10^{-5}$) |
| SiC/T | 2.21 | 2.1 | 23.6 | 21.5 ($10^{-7}$) | 18.9 ($10^{-8}$) | 14.6 ($10^{-9}$) |
| SiC/Pm-147 | 2.33 | 2.30 | 24.9 | 5.3 ($10^{-7}$) | 1.9 ($10^{-8}$) | 0.65 ($10^{-8}$) |

Table 3. The limiting and achievable parameters of a betavoltaic converter

The, attainable and limiting betaconversion efficiencies calculated by (13) and (14) using Si and SiC parameters are shown in Table 3.

The obtainable efficiencies $\eta_{real}$ are calculated using collection coefficients $Q$ from Figs. 1 and 2. The lifetimes of $10^{-3}$, $10^{-4}$ and $5 \cdot 10^{-5}$ s were used for Si, while for SiC the lifetimes of $10^{-7}$, $10^{-8}$ and $10^{-9}$ s were used. The table shows that $\eta_{real}$ decreases slightly with the decrease of $\tau_{SR}$. $\eta_{real}$ is significantly lower for SiC with Pm-147 as the source of electrons than for the three remaining source/material combinations. The similar situation is expected for direct gap semiconductors with lifetimes of $10^{-7} - 10^{-9}$ s. The reason is the change of the $\alpha L > 1$ inequality needed for the full electron-hole collection to the $\alpha L < 1$ inequality.

Note that the experimental efficiency is 4% for Si/Pm-147 system with two-sided irradiation [11]. The attainable $\eta$ value for Si/Pm-147 system is 13.6 % (see Table 1).

## *4. Conclusions*

We develop a formalism to determine the maximum attainable conversion efficiency $\eta_{lim}$ in the Si and SiC based betavoltaics batteries, using [147]Pm and tritium as the source of beta-electrons. The formalism of carrier transport and collection is based on the similarity of betavoltaic and photovoltaic processes, however, several important differences have been included in the description of the physical processes in the beta-voltaics systems. The realistic experimentally achieved parameters of the betavoltaic systems were included in the analytical formalism developed. This allows us to also calculate the realistically achievable betavoltaic efficiencies.

We calculate the collection coefficient achievable $Q$ and open-circuit voltage $V_{OC}$. The betavoltaic conversion efficiency obtained for high-quality Si $p-n$ junction is higher than for SiC $p-n$ junction with $Pm-147$ source. This is due to the high electron-hole pairs' generation depth and low diffusion length in SiC. Therefore, the majority of the generated electron-hole pairs recombine and do not reach the $p-n$ junction. The situation for direct-gap semiconductors (particularly GaAs) with low minority carriers lifetime and low diffusion length is similar.

Tritium with low beta-electron energy and low pairs generation depth is a better source for SiC with low diffusion length. The collection coefficient for this case is high enough. Therefore, SiC/T conversion efficiency can be greater than Si/T efficiency because of the higher SiC band gap, and consequently higher open circuit voltage.

The influence of the dead layer on the collection efficiency value is also included in the formalism and analyzed. The dead layer is associated with a nonmonotonic electron-hole pair generation function in the semiconductor under electrons flow irradiation. The dependence of the collection coefficient $Q$ on the dead layer thickness $x_m$ is established. This dependence is weak for the case of diffusion length $L$ significantly greater than $x_m$ value.

The calculated efficiencies indicate a limit to the maximum possible performance of the betavoltaic systems, e.g., $\eta = 8\%$ for the tritium based $^3$H/Si system. While being comparable to experimentally achieved efficiencies, our results demonstrate that there is still sufficient room for efficiency increase using optimized materials parameters and the system design.

# References

1. L.C. Olsen, P. Cabauy, and B.J. Elkind, Betavoltaic power sources, Physics Today, December, 2012, p. 35.

2. L.C. Olsen, "Review of betavoltaic energy conversion", NASA TechDoc 19940006935, http://hdl.handle.net/2060/19940006935

3. V.M. Andreev, A.G. Kevetsky, V.S. Kaiinovsky, V.P. Khvostikov, V.R. Larionov, V.D. Rumyantsev, M.Z. Shvarts, E.V. Yakimova, V.A. Ustinov, Tritium-powered betacells based on $Al_xGa_{1-x}As$, Conference Record of the Twenty-Eighth IEEE Photovoltaic Specialists Conference -2000, IEEE, Piscataway, NJ (2000), pp.1253-1256, DOI: 10.1109/PVSC.2000.916117

4. K.E. Bower, Y.A. Barbanel, Y.G. Shreter, G.W. Bohnert, Polymers, Phosphors, and Voltaics for Radioisotope Microbatteries, CRC Press, Boca Raton, Fl (2002), p.46

5. T.Adams, S. Revankar, P. Cabauy, L.C. Olsen, Status of Betavoltaic Power Sources for Nano and Micro Power Application, 45th Power Sources Conference, Las Vegas, Nevada, USA, 11 - 14 June 2012, pp.318-321

6. W. Ehrenberg, Chi-Shi Lang and R. West, The Electron Voltaic Effect, Proc. Phys. Soc. A **64**, p.424 (1951). doi:10.1088/0370-1298/64/4/109

7. P.Rappaport, The Electron-Voltaic Effect in p-n Junctions Induced by Beta Particle Bombardment, Phys. Rev. **23**, pp.246-247 (1953). http://dx.doi.org/10.1103/PhysRev.93.246.2

8. W.G. Pfan and W. Van Roosbroeck, Radioactive and Photoelectric p-n Junction Power Sources, J.Appl. Phys. **25**, pp.1422-1434 (1954). DOI: 10.1063/1.1721579

9. P. Rappaport, J. J. Loferski and E. G. Linder, The Electron-Voltaic Effect in Germanium and Silicon P-N Junctions, RCA Rev. **17,** p.100 (1956).

10. H. Flicker, J. J. Loferski and T. S. Elleman, Construction of a Promethium-147 Atomic Battery, IEEE Trans, ED-**11**, pp.2-8 (1964). DOI: 10.1109/T-ED.1964.15271

11. L. C. Olsen, Betavoltaic Energy Conversion, Energy Conversion **12**, pp.117-124 (1973).

12. L.C. Olsen, Advanced Betavoltaic Power Sources, Proc. 9th Intersociety Energy Conversion Engineering Conference, pp.754-762 (1974).

13. A.L. Fahrenbruch and R.H. Bube. Fundamentals of solar cells. Photovoltaic solar energy conversion. Academic Press, New York, 1983

14. Kyuhak Oh, M.A. Prelas, J.B. Rothenberger, E.D. Lukosi, J. Jeong, D.E. Montenegro, R.J. Schott, C.L. Weaver and D.A. Wisniewski, Theoretical Maximum Efficiency of Optimized Slab and Spherical Betavoltaic Systems Utilizing Sulfur-35, Strontium-90, and Yttrium-90, Nuclear Technology **179**, pp.234-242 (2012).

15. S. Mertens, T. Lasserre, S. Groh, F. Glueck, A. Huber, A. W. P. Poon, M. Steidl, N. Steinbrink, C. Weinheimer, Sensitivity of Next-Generation Tritium Beta-Decay Experiments for keV-Scale Sterile Neutrinos, http://arxiv.org/abs/1409.0920

16. T. Baltakmens, Accuracy of absorption methods in the identification of beta emitters, Nuclear Instruments and Methods, **142**, pp.535-538 (1977).

17. J.M. Trischuk, Nuclear beta spectroscopy using solid state detectors, PhD thesis, California Institute of Technology, Pasadena, California, 1967.

18. C.A. Klein, Bandgap Dependence and Related Features of Radiation Ionization Energies in Semiconductors, J. Appl. Phys., **39**, pp.2029-2038 (1968). DOI: 10.1063/1.1656484

19. N.L. Dmitruk, V.G. Litovchenko and G.H. Talat, The effect of the surface space charge region on the cathodoluminescence of semiconductors, Surf. Sci, **72**, pp.321-341 (1978). doi:10.1016/0039-6028(78)90298-4

20. A.P. Gorban, A.V. Sachenko, V.P. Kostylyov and N.A. Prima, Effect of excitons on photoconversion efficiency in the $p^+$-n-$n^+$- and $n^+$-p-$p^+$-structures based on single-crystalline silicon, Semiconductor Physics, Quantum Electronics and Optoelectronics, **3**, pp.322-329 (2000).

21. H.C. Casey, Jr. and R.H. Kaiser, Analysis of N‐Type GaAs with Electron‐Beam‐Excited Radiative Recombination, J. Electrochem. Soc. **114**, pp.149-153 (1967). DOI: 10.1149/1.2426527

22. A. Galeskas, J. Linnros and B. Breitholtz, Time-resolved imaging of radiative recombination in 4H-SiC p-i-n diodes, Appl. Phys. Lett., **74**, 3398-3400 (1999). http://dx.doi.org/10.1063/1.123363

23. W. Shockley and H.J. Queisser, Detailed balance limit of efficiency of *p-n* junction solar cells, J. Appl. Phys. **32**, pp.510-519 (1961). DOI: 10.1063/1.1736034